\documentstyle[prd,aps]{revtex}
\begin{document}
\title{High Temperature Asymptotics in Terms of Heat Kernel Coefficients:
Boundary Conditions with Spherical and Cylindrical Symmetries}
\draft
\author{M. Bordag\thanks{E-mail: Michael.Bordag@itp.uni-leipzig.de}}
\address{Universit\"at Leipzig, Institut f\"ur Theoretische Physik \\
Augustusplatz 10, 04109 Leipzig, Germany}
\author{V.V. Nesterenko\thanks{E-mail: nestr@thsun1.jinr.ru},
I.G. Pirozhenko\thanks{E-mail: pirozhen@thsun1.jinr.ru}}
\address{Bogoliubov Laboratory of Theoretical Physics \\
Joint Institute for Nuclear Research, Dubna, 141980, Russia}
\date{\today}
\maketitle

\begin{abstract}
     The high temperature asymptotics of the Helmholtz free energy of
electromagnetic field subjected to boundary conditions with spherical
and cylindrical symmetries are constructed by making use of a general
expansion in terms of heat kernel coefficients and the related
determinant.  For this, some new heat kernel coefficients and
determinants had to be calculated for the boundary conditions under
consideration.  The obtained results reproduce all the asymptotics
derived by other methods in the problems at hand and involve a few new terms
in the
high temperature expansions. An obvious merit of this approach is its
universality and applicability to any boundary value problem correctly
formulated.
\end{abstract}

\vspace{0.5cm}
The Casimir calculations at finite temperature prove to be a nontrivial problem
specifically
for boundary conditions with nonzero curvature.
 Investigation of the high temperature limit,
i.e.\ the classical limit, in this problem is of
independent interest~\cite{FMR}. For this goal
a powerful method of the zeta function technique and the heat kernel
expansion can be used. It is essential that for obtaining
the high temperature  asymptotics of the thermodynamic
characteristics it is sufficient to know
the heat kernel coefficients and the determinant   for the
spatial  part of the  operator governing the field
dynamics. This determinant
is the derivative of its zeta function in zero,
whereas the Casimir energy at zero temperature
is given by this zeta function in the point $-\frac12$.
 This is an essential merit
of this approach.
     This high temperature expansion in terms of the heat kernel
coefficients and the zeta
function determinant is known for a long time~\cite{DK},
but did not receive the due attention
and has to our knowledge not been used in the case of boundary conditions.
 The present study seeks to fill the gap in this
area and to demonstrate the universality and efficiency of this
approach by treating a number of nontrivial boundary conditions for
the electromagnetic field.
We present here only the basic results of our calculations.

Starting point in our consideration is the general high temperature expansion of the
free energy initially given in \cite{DK} and recently discussed in
\cite{quant-ph/0106045}, Sec. 5.1.3,
\begin{eqnarray}
F(T)&\simeq&-\frac{T}{2}\zeta'(0)+a_0\frac{T^4}{\hbar^3}\,
\frac{\pi^2}{90}-\frac{a_{1/2}T^3}{4\pi^{3/2} \hbar^2 }
\zeta_R(3) -\frac{a_1}{24}\frac{T^2}{\hbar}
+\frac{a_{3/2}}{(4\pi)^{3/2}}T\ln\frac{\hbar}{T}
-\frac{a_2}{16\pi^2}\hbar \left[\ln\left(\frac{\hbar}{4
\pi
T}\right)+\gamma\right]\nonumber\\&&-\frac{a_{5/2}}{(4\pi)^{3/2}}\frac{\hbar^2}{24
T}-T\sum_{n\geq
3}\frac{a_n}{(4\pi)^{3/2}}\left(\frac{\hbar}{2 \pi
T}\right)^{2n-3}
\,
 \Gamma(n-3/2)\,\zeta_R(2 n-3)\,{.}
 \label{F}
\end{eqnarray}
Here $\gamma$ is the Euler constant and $\zeta_{R}(s)$ is the Riemann
zeta function. The quantities under the
logarithm sign in expansion (\ref{F}) are dimensional, but upon collecting
similar terms with account for the logarithmic ones in
$\zeta'(0)$ it is easy to see that  finally the  logarithm function
has a
dimensionless argument.

 It is worth noting that the zeta function determinant,
i.e., $\zeta'(0)$, does not enter the asymptotic
expansion for the internal energy which is completely  defined only by the heat
kernel coefficients. In view of this, the first term in the asymptotics
of the free energy in Eq.\ (\ref{F}) is referred to as a pure entropic
contribution~\cite{FMR}. Its physical origin is  till now not elucidated.
The entropic term is a pure
classical quantity because it does not  involve the Planck
constant $\hbar$.  This classical contribution to the asymptotics seems
 to be derivable
without appealing to the notion of quantized electromagnetic field.

The heat kernel
coefficients needed for construction of the expansion (\ref{F})
will been  calculated as the residua of the
corresponding zeta functions. For the boundary conditions under
consideration the explicit expressions for the zeta functions have been
derived in our papers~\cite{LNB,BP}. The zeta function determinants
involved in Eq.\ (\ref{F}) are calculated in Ref.~\cite{BNP} using the methods
of Ref. \cite{BGKE}. For a simple calculation of the relevant heat kernel
coefficients see also Ref. \cite{BKE}

{\it A perfectly conducting spherical shell of radius $R$ in vacuum.}
The first six heat kernel coefficients in this problem are:
\begin{eqnarray}
a_0=0, \quad a_{1/2}=0,\quad a_1=0, \quad \frac{a_{3/2}}{(4\pi)^{3/2}}=
\frac{1}{4},
\quad
a_2=0,\quad \frac{a_{5/2}}{(4\pi)^{3/2}}=\frac{c^2}{160\,R^2}.
\label{eq_2}
\end{eqnarray}
Furthermore
\begin{equation}
a_j=0,\qquad j=3,4,5,  \dots\, {.}
\label{eq3}
\end{equation}
 The exact value of  $\zeta' (0)$ is
derived in~\cite{BNP}
\begin{equation}
\zeta' (0)=\left (\frac{1}{2}-\frac{\gamma}{2}+\frac{7}{6}\ln
2+6\,\zeta'_R(-1)   \right )
+\left(-\frac{5}{8}+\frac{1}{2}\,\ln\frac{R}{c}+\ln
2+\frac{\gamma}{2} \right)
=0.38429+\frac{1}{2}\ln\frac{R}{c}.
\label{eq4}
\end{equation}
As a result we have the following high temperature asymptotics
of the  free energy
\begin{equation}
F (T)=-\frac{T}{4}\left({\displaystyle 0.76858}+\ln \tau+
 \frac{1}{960\tau ^2}\right )+{\cal O }(T^{-3}),
\label{eq5}
\end{equation}
where $\tau =RT/(\hbar c)$ is the dimensionless `temperature'.
The expression (\ref{eq5}) exactly reproduces the  asymptotics
obtained in Ref.\ \cite{BD} by making use of the multiple scattering
technique (see Eq.~(8.39) in that paper).

{\it A compact ball with $c_1=c_2$.}
In this case the spherical surface  delimits the media with
``relativistic invariant'' characteristics i.e., the velocity of light is
the same
inside and
outside the ball~\cite{Brevik}. Here there naturally  arises
a dimensionless parameter~\cite{BrNP}
\begin{equation}
\xi^2=\left(\frac{\varepsilon_1-
\varepsilon_2}{\varepsilon_1+\varepsilon_2}\right)^2=
\left(\frac{\mu_1-\mu_2}{\mu_1+\mu_2}\right)^2,
\end{equation}
where $\varepsilon_1$ and $\varepsilon_2$ ($\mu_1$ and $\mu_2$)
are permittivities (permeabilities) inside and outside the
ball. As usual we perform the calculation in the first order
of the expansion with respect to $\xi^2$.

The zeta function for this boundaries, obtained in Ref.\ \cite{LNB},
affords the exact values of heat kernel
coefficients up to $a_3$
\begin{equation}
a_0= a_{1/2}= a_1=0,\quad
a_{3/2}=2\pi^{3/2}\xi^2,  \quad
 a_2=0, \quad  a_{5/2}=0, \quad
a_3=0.\label{eq4_14}
\end{equation}
The zeta determinant in this problem
turns out to be given  by multiplication of the content of the second parentheses
in Eq.\ (\ref{eq4})  by $\xi^2$
\begin{equation}
\zeta'(0)
=\xi^2\left(0.35676+\frac{1}{2}\ln\frac{R}{c}\right).
\label{eqbz}
\end{equation}
The high temperature asymptotics for free energy reads

\begin{equation}
F(T)=-\xi^2\frac{T}{4}\left(\gamma+\ln4-\frac{5}{4}+
\ln\tau \right )+{\cal O}(T^{-3})
=-\xi^2\frac{T}{4}\left (0.71352+
\ln \tau \right )+{\cal O}(T^{-3}){.}
\label{eq4_17}
\end{equation}
The asymptotics (\ref{eq4_17})  completely
coincide with analogous formula obtained in Refs.\ \cite{NLS,KFMR} by
the mode summation method combined with the addition theorem for
the Bessel functions.

{\it A pure dielectric ball.}
In this case the heat kernel coefficients $a_0$ and $a_{1/2}$ prove to be
nonvanishing \cite{BKV}. Their contributions to the high temperature asymptotics
should be removed because the Stefan-Boltzmann constant should not be
renormalized. For the next coefficients we have
\begin{equation}
a_1=a_2=0, \quad a_{3/2}=\pi^{3/2}\Delta n^2, \quad a_{5/2}=0,
\label{eq4_20}
\end{equation}
where
 $\Delta
n=n_1-n_2=n_1\,n_2\,(c_2-c_1)/c\simeq(c_2-c_1)/c$, $n_i$ and
$c_i$ are the  refractive index and  the velocity of light inside ($i=1$)
and outside ($i=2$) the ball, and $c$ is the velocity of light in
the vacuum. It is assumed that  $c_1$ and $c_2$
differ from $c$ slightly, therefore $c_2-c_1$ and $\Delta n$ are small
quantities.
Making use of the technique developed  in Ref.\ \cite{BGKE} one
obtains the following expression for the
zeta function determinant~\cite{BNP}
\begin{equation}
\zeta'(0)=\frac{\Delta
n^2}{4}\left(-\frac{7}{8}+\ln\,\frac{R}{c}+\ln 4 + \gamma
\right). \label{eq4_21}
\end{equation}
The high temperature asymptotics of free energy reads
\begin{equation}
F(T)=-\frac{\Delta n^2}{8}T\left ( \ln 4\tau +\gamma -\frac{7}{8}
\right ) +{\cal O}(T^{-2}). \label{fdb}
\end{equation}
It is worth comparing  these results with
analogous asymptotics obtained by different methods.
In Ref.\ \cite{NLS}
at the beginning of calculations the first term of
expansion of the internal
energy  was derived. The subsequent integration of the
thermodynamic relation  gave the correct
coefficient of the logarithmic term in the asymptotics of free energy
(\ref{fdb}).
Barton~\cite{Barton} managed to deduce the asymptotics
(\ref{fdb}) except for the contributions
proportional to negative powers of $T$. One should keep in mind
that  our parameter $\Delta n$ corresponds to
$2\pi\alpha\, n$ in the notations of Ref.~\cite{Barton}.
The asymptotics
(\ref{fdb}) contain  the $R$-independent terms.
As far as we know the
physical meaning of such  terms remains
unclear.

{\it A perfectly conducting cylindrical shell.}
The heat kernel coefficients are
\begin{equation}
a_0= a_{1/2}= a_1= a_2=0,\quad
\frac{a_{3/2}}{(4\pi)^{3/2}}=\frac{3}{64\,R}, \quad
\frac{a_{5/2}}{(4\pi)^{3/2}}=\frac{153}{8192}\frac{c^2}{R^3}.
\end{equation}
The zeta function determinant in this problem is calculated in~\cite{BNP}
\begin{equation}
\zeta'(0)=\frac{0.45711}{R}+\frac{3}{32\,R}\,\ln\frac{R}{2\,c}.
\label{eq5_6}
\end{equation}
The free energy behavior at high temperature is the following
\begin{equation}
F(T)=-\frac{T}{R} \left (0.22856 +\frac{3}{64}
\ln\frac{\tau }{2}-\frac{51}{65536\tau ^2}\right )
+{\cal O}(T^{-3}).
\label{fcs}
\end{equation}
 The high temperature asymptotics  of the electromagnetic
free energy in presence of perfectly conducting cylindrical
shell was investigated  in Ref.\ \cite{BD}. To make the
comparison handy let us rewrite  their result as follows
\begin{equation}
F(T)\simeq-\frac{T}{R}\left (0.10362+\frac{3}{64R}
\ln\frac{\tau}{2}\right ). \label{15a}
\end{equation}
The discrepancy between the  terms linear  in $T$
in Eqs.\
(\ref{fcs}) and (\ref{15a}) is due to the double scattering
approximation used in Ref.\ \cite{BD} (see also below).
Our approach provides  an opportunity to calculate the exact value of
this term (see Eq.\  (\ref{fcs})).

{\it A compact infinite cylinder with $c_1=c_2$.}
By making use of the relevant zeta function  found in Ref.\ \cite{LNB}
one obtains the heat kernel coefficients
\begin{eqnarray}
a_0= a_{1/2}= a_{1}=a_2= a_j=0,\;j=3,4, \ldots ,
\nonumber \\   \frac{a_{3/2}}{(4\pi)^{3/2}}
=\frac{3\xi^2}{64\,R}, \;\;
\frac{a_{5/2}}{(4\pi)^{3/2}}=\xi^2\frac{c^2}{R^3}\frac{45}{8192}
{.}
\label{ceh}
\end{eqnarray}
The corresponding zeta function determinant is calculated in Ref.\ \cite{BNP}
\begin{equation}
\zeta'(0)=
\frac{\xi^2}{R}\left( 0.20483+
\frac{3}{32}\,\ln\frac{R}{2\,c}\right).
\label{eq5_14}
\end{equation}
Now we can write the high temperature asymptotics for free energy
\begin{equation}
F(T)= -\xi^2\frac{T}{R}\left (0.10242 -\xi^2\frac{3}{64}\ln \tau
+\frac{15}{65536\tau ^2} \right )
+{\cal O}(T^{-3}) .
\label{eq18}
\end{equation}
  Putting in Eq.\ (\ref{eq18}) $\xi^2=1$ we arrive at the double scattering
approximation for a perfectly conducting cylindrical shell (\ref{15a}).

{\it A pure dielectric cylinder.}
In this problem the first three heat kernel coefficients prove to
be nonzero~\cite{BP}. Their
contributions to the high temperature asymptotics should be canceled by respective
redefinition of the parameters in the general expression for free energy of such a
cylinder (in the same way as  it has been done
for a pure dielectric ball).
The next coefficients are given by
\begin{equation}
 a_2=0,  \quad
 a_{3/2}=\frac{3\pi^{3/2}}{16 R c_2^2}\,(c_1-c_2)^2,
\quad
\frac{a_{5/2}}{(4\pi)^{3/2}}=
\frac{857}{61440}\frac{(c_1-c_2)^2}{R^3} \label{eq19}.
\end{equation}
On this basis we can write the  high temperature asymptotics
for internal energy in the problem at hand
\begin{equation}
U(T)=  \Delta n^2  \frac{3}{128}\frac{T}{R}\left ( 1-
\frac{857}{17280\tau ^2} \right ) +{\cal O}(T^{-2}){,}
\label{eq20}
\end{equation}
where $\Delta n=n_1-n_2\simeq(c_2-c_1)/c$.
In view of  considerable technical difficulties we shall not
calculate the zeta function determinant for a pure dielectric
cylinder. We recover the respective asymptotics for free energy by
integrating the appropriate  thermodynamic relation using Eq.\
 (\ref{eq20}). Pursuing this way we introduce
a new constant of integration $\alpha $ that remaines undetermined
\begin{equation}
F(T)=-\Delta n^2\frac{3 }{128}\frac{T}{R}\left (\alpha + \ln \tau
 +\frac{857}{34560\tau ^2}
\right )+{\cal O}(T^{-2})\,{.}
\end{equation}

To sum up we have demonstrated efficiency and universality of the
high temperature expansions in terms of the heat kernel coefficients
for the Casimir problems with spherical and cylindrical symmetries. All
the known results in this field are reproduced  in a uniform approach
and in addition  a few new asymptotics are derived (an exact asymptotics for a
perfectly conducting cylindrical shell, for a compact cylinder
with $c_1=c_2$, and  for a pure dielectric infinite cylinder).

As the next step in the development of this approach one can try to
retain  the terms exponentially  decreasing when $T\to \infty $.
These corrections are well known, for example, for thermodynamic
functions of electromagnetic field in the presence of perfectly
conducting parallel plates \cite{PMG}. In order to reveal  such
terms, first of all the exponentially decreasing corrections should be
retained in the standard asymptotic expansion for the heat kernel.

As in all the Casimir calculations taking into account the material
characteristics  of the boundaries, in the approach under
consideration the renormalization procedure  should be formulated
exactly in the framework of the high temperature expansions
(\ref{F}).
The explicit
divergencies are not encountered here, but
nevertheless the renormalization should be carried out as it was
shown  in this paper when considering a pure dielectric ball and
dielectric cylinder.

\end{document}